\documentstyle[aps,prl]{revtex} 

\begin{document}


\title{Field Redefinition Invariance in Quantum Field Theory}
                                                                               
\author{Karyn M. Apfeldorf,$^{\, a}$
Horacio E. Camblong,$^{b}$
and
Carlos R. Ord\'o\~nez$^{c,d}$}

\address{$^a$ Aret\'{e} Associates, P.O. Box 6024,
Sherman Oaks, CA 91413 \\
$^b$ Department of Physics, University of San Francisco, San
Francisco, CA 94117-1080 \\
$^c$ Department of Physics, University of Houston, Houston,
TX 77204-5506
\\
$^d$
World Laboratory Center for Pan-American Collaboration in Science and
Technology 
}
                                               
\maketitle

\begin{abstract}
The issue of field redefinition 
invariance 
of path integrals in quantum field theory is reexamined.
A ``paradox'' is presented involving the reduction 
 to an effective quantum-mechanical theory of 
a $(d+1)$-dimensional free scalar field
in a Minkowskian spacetime with compactified spatial coordinates.
The implementation of field redefinitions both before and after 
the reduction suggests that operator-ordering 
issues in quantum field theory should not be ignored.  
\end{abstract}




\def\diagram#1{{\normallineskip=8pt
       \normalbaselineskip=0pt \matrix{#1}}}

\def\diagramrightarrow#1#2{\smash{\mathop{\hbox to 1.2in{\rightarrowfill}}
        \limits^{\scriptstyle #1}_{\scriptstyle #2}}}

\def\diagramleftarrow#1#2{\smash{\mathop{\hbox to 1.2in{\leftarrowfill}}
        \limits^{\scriptstyle #1}_{\scriptstyle #2}}}

\def\diagramdownarrow#1#2{\llap{$\scriptstyle #1$}\left\downarrow
    \vcenter to .8in{}\right.\rlap{$\scriptstyle #2$}}

\def\diagramuparrow#1#2{\llap{$\scriptstyle #1$}\left\uparrow
    \vcenter to .8in{}\right.\rlap{$\scriptstyle #2$}}

\def\fo{\hbox{{1}\kern-.25em\hbox{l}}}
\def\fnote#1#2{\begingroup\def\thefootnote{#1}\footnote{#2}\addtocounter
{footnote}{-1}\endgroup}
\def\ul{\underline}
\def\indt{\parindent2.5em}
\def\ch{\@startsection{section}{1}{\z@}{-3ex plus-1ex minus-.2ex}%
        {2ex plus.2ex}{\large\sc}}
\def\7#1#2{\mathop{\null#2}\limits^{#1}}        
\def\5#1#2{\mathop{\null#2}\limits_{#1}}        
\def\too#1{\stackrel{#1}{\to}}
\def\tooo#1{\stackrel{#1}{\longleftarrow}}
\def\dddot#1{\hbox{$\mathop{#1}\limits^{\ldots}$}}
\def\ddddot#1{\mathop{#1}\limits^{\ldotp\ldotp\ldotp\ldotp}}
 
%
\newcommand{\NP}[3]{{Nucl. Phys.} (19#2)  {#3}}
\newcommand{\NPB}[3]{{Nucl. Phys.} {\bf B#1} (19#2)  {#3}}
\newcommand{\JPA}[3]{{J. Phys.} {\bf A#1} (19#2)  {#3}}
\newcommand{\PRD}[3]{{Phys. Rev.} {\bf D#1} (19#2)   {#3}}
\newcommand{\PLB}[3]{{Phys. Lett.} {\bf #1B} (19#2)  {#3}}
\newcommand{\PRL}[3]{{Phys. Rev. Lett.} {\bf #1} (19#2) {#3}}
\newcommand{\ZFP}[3]{{Zeitsch. f. Physik} {\bf C#1} (19#2)  {#3}}
\newcommand{\PR}[3]{{Phys. Rev.} {\bf #1} (19#2)  {#3}}
\newcommand{\AP}[3]{{Ann. of Phys. (N.Y.)} {\bf #1} (19#2)  {#3}}
\newcommand{\IJA}[3]{{Int. Journ. of Mod. Phys.} {\bf A#1} (19#2) {#3}}
\newcommand{\CMP}[3]{{Commun. Math. Phys.} {\bf #1} (19#2) {#3}}
\newcommand{\JMP}[3]{{J. Math. Phys.} {\bf #1} (19#2) {#3}}
\newcommand{\LMP}[3]{{Lett. Math. Phys.} {\bf #1} (19#2) {#3}}
\newcommand{\MPLA}[3]{{Mod. Phys. Lett.} {\bf A#1} (19#2) {#3}}
\newcommand{\PRep}[3]{{Phys. Rep.} {\bf #1} (19#2)  {#3}}
\newcommand{\CQG}[3]{{Class. Quant. Grav.} {\bf #1} (19#2)  {#3}}
\newcommand{\PTP}[3]{{Prog. Theor. Phys.} {\bf #1} (19#2)  {#3}}
\newcommand{\TMP}[3]{{Th. Math. Phys.} {\bf #1} (19#2)  {#3}}
\newcommand{\ZPC}[3]{{Z. Phys.} {\bf C#1} (19#2)  {#3}}
\newcommand{\PRS}[3]{{Proc. Roy. Soc.} {\bf A#1} (19#2)  {#3}}

\def\tr{{\rm Tr}}
\def\dgg{\dagger}
\def\ddt#1{\frac{d \;}{dt} (#1)}

\def\beq{\begin{equation}}
\def\eeq{\end{equation}}
\def\beqar{\begin{eqnarray}}
\def\eeqar{\end{eqnarray}}
\def\beqarn{\begin{eqnarray*}}
\def\eeqarn{\end{eqnarray*}}
\def\half{\frac{1}{2}}

Field redefinition 
invariance is a basic property expected of
all physically meaningful quantities,
such as the poles of renormalized  
propagators.
By contrast, there exist quantities related to the
specific choice of field variables, such as
wave function renormalization factors, whose
values depend on the particular parametrization.

The question then arises as to the transformation properties of 
functional integrals under nonlinear point canonical transformations.
For the quantum-mechanical counterpart of this problem, 
additional terms $O(\hbar^2)$ appear in the path 
integral~\cite{dew:57,edw:64,gro:98}.
This phenomenon---which can be viewed as a manifestation of the 
stochastic nature of the Lagrangian formulation of the
path integral---is usually studied by introducing a
discretization of the time variable~\cite{mcl:71,ger:76,apf:96}. 
The ensuing ``extra'' terms are an inevitable
consequence of the quantization of the theory, which
promotes the classical coordinates to quantum operators;
in effect,
for every operator ordering of the associated 
Hamiltonian,\footnote{In this Letter we adopt
Weyl ordering, which corresponds
to the midpoint prescription.}
 there exists a particular
prescription for handling the lattice definition of the
path integral~\cite{gro:98}.

The standard lore in quantum field theory dictates,
in contradistinction to the quantum-mechanical procedure,
that no additional terms are needed. 
More precisely,
the action is assumed to change by direct 
substitution of the field transformation, together with the 
inclusion of a term arising from the Jacobian determinant associated with the 
change of field variables~\cite{chi:61}.  
Moreover, for a $D$-dimensional quantum
field theory, the Jacobian becomes superfluous
within the dimensional-regularization scheme---upon exponentiation,
the formally infinite $D$-dimensional
spacetime delta function $\delta^{(D)} (0)$ 
generated by the trace is set equal to zero~\cite{diagrammar}.  
A similar line of reasoning is employed to argue away any possible
contribution by ``extra'' terms generated in the path integral; 
these terms---being a manifestation of operator ordering---would 
vanish by dimensional regularization, because they
would involve delta functions at zero spatial argument, as follows
from $ [\Phi (x), \Pi (x) ]  = i \hbar \delta^{(D-1)}(0)$.
More precisely,
the standard justification for this procedure
is based on the assumed existence and necessity of local counterterms in the
action, so that Jacobians and any other additional
terms resulting from operator 
ordering (all of which are local quantities in the action), 
have the only effect of changing the coefficients of these local counterterms.

However, upon closer examination, 
one realizes that a solid justification for setting infinite quantities 
equal to zero is still lacking.
Even if the validity of dimensional regularization is not questioned, 
one could analyze the problem from the lattice point of 
view,\footnote{The spacetime delta function at zero argument 
is related to the inverse lattice spacing $a$,
in the form  $\delta^{(D)}(0) \sim  a^{-D}$.} 
in which the Jacobian as well as the ``extra'' terms, 
do not vanish.
In fact, the relevance of a term proportional
to $\delta^{(D)}(0)$,
which may be interpreted as a limitation of dimensional
regularization,
was discovered in the early literature of the massive vector
boson theory~\cite{lee:62,sal:70}
and of the renormalization of the nonlinear sigma 
model~\cite{sal:70}, where 
it was used for the explicit cancellation of divergent 
terms~\cite{hon:71,cha:71,ger:71}.

The purpose of this Letter is to investigate 
these questions in a field theory toy model, in which we have full control 
of regularization issues
and can test the relevance of non-linear field redefinitions. 
Further technical details will appear elsewhere.
Our model 
is a free scalar quantum field
theory in $D=d+1$ dimensions, in a 
flat spacetime with Minkowskian metric 
$\eta^{\mu \nu} ={\rm diag} (+1,-1, \ldots, -1)$,
characterized by the action
\begin{eqnarray}
S [  \Phi ]
& = & \half \int_{ \Re^{1} \times T^{d}}
d^{D} x \; 
\left(
\eta^{\mu \nu} \partial_\mu \Phi  \,
\partial_\nu \Phi
-
m^{2} \Phi^{2}
\right)
\nonumber \\
 & = &
\frac{1}{2}  \int_{ \Re^{1}} d t \; \int_{ T^{d}} 
d^{d} {\bf x} \;
\left\{
\left[
\dot{\Phi}(t,{\bf x})
\right]^{2} 
-
\left[
 \mbox{\boldmath $\nabla$} \Phi (t,{\bf x})
\right]^{2} 
- m^{2}
\left[
 \Phi (t,{\bf x})
\right]^{2} 
\right\} 
\;  .
\label{eq:action_continuous}
\end{eqnarray}
In Eq.~(\ref{eq:action_continuous}),
each spatial coordinate (corresponding to $\mu=1,\ldots, d$)
is assumed to be curled up into a circle $S^{1}$ of radius $R$,
so that the whole space is compactified into a $d$-dimensional torus 
$T^{d}=S^{1} \times \ldots \times S^{1}$;
this amounts to the periodicity conditions
$x^{\mu} \cong x^{\mu} + L$ (for $\mu=1,\ldots, d$),
with $L=2 \pi R  $, 
which permit a simplification in our analysis of field 
redefinitions.\footnote{Our selection of the compact space $T^{d}$ 
is guided by the convenience of choosing a flat spacetime.
Our analysis suggests that the ``extra terms'' will arise 
independently from the details of this compactification procedure.}
Our approach is based on making a nonlinear but local field redefinition 
\beq
\Phi = F[\tilde{\Phi}] 
\;  ,
\label{eq:Ntran}
\eeq
using two different methods
and comparing the corresponding results.
In Method 1,
the compactified spatial coordinates are integrated out to yield an
effective quantum-mechanical problem, which is then subject
to the quantum-mechanical counterpart of the transformation~(\ref{eq:Ntran}),
for which
the existence of an ``extra'' term is a
well-established result~\cite{apf:96}. 
In Method 2, 
the standard quantum-field theoretical lore 
is applied directly to the field redefinition~(\ref{eq:Ntran})
of the $D$-dimensional field theory, followed
by a reduction  to a quantum-mechanical action by 
integrating out the spatial coordinates.
The required identity of the results of the two methods
leads to a remarkable ``paradox'':
no new terms are developed
in the field-theory case (Method 2), despite
the appearance of ``extra'' terms 
for the quantum-mechanical case (Method 1).
Reconciling these two methods 
calls for either a detailed explanation
or a revision of the standard lore.
   
A digression
is in order for subsequent notational and computational purposes.
In our Letter we will exploit the local nature of the field
redefinition~(\ref{eq:Ntran}), which guarantees the 
ultralocal property of the spacetime metric, i.e.,
\begin{equation}
{\cal G} [ \tilde{\Phi} ] (t,{\bf x};t^{\prime},{\bf x}^{\prime})
=
\delta (t-t^{\prime}) \, 
\delta^{(d)} ({\bf x}-{\bf x}^{\prime})
\,
\left(
 F^{\prime}[\tilde{\Phi} (t,{\bf x})] 
\right)^{2}
= \delta (t-t^{\prime}) \,  g [\tilde{\Phi} ] 
 ({\bf x}, {\bf x}^{\prime}; t)
\;  ,
\label{eq:ultralocal_metric}
\end{equation}
where the reduced metric  $ g [\tilde{\Phi} ]$ will be 
useful for the calculations of Method 1.
When implementing the nonlinear field redefinition~(\ref{eq:Ntran}),
 the transformed action $S[\tilde{\Phi}]$
 has two (Method 2) or three (Method 1) pieces:
(i)
the part obtained by direct substitution into 
the original free action~(\ref{eq:action_continuous}),
  $S_{0} [\tilde{\Phi} ] 
= S \left[ F[\tilde{\Phi}] \right] $  
;
(ii) 
the effective action arising from the Jacobian,
\begin{equation}
S_{\rm Jacobian}[\tilde{\Phi}]= 
- \frac{i \hbar}{2} \;
{\rm Tr} \, \ln {\cal G} [\tilde{\Phi}]
\;  ,
\label{eq:jacobian_spacetime}
\end{equation}
where ${\rm Tr}$ stands for the spacetime trace;
 and (iii)
the ``extra'' term
$S_{\rm extra} [\tilde{\Phi}] $  of $O(\hbar^2)$
arising from
its quantum-mechanical Weyl-ordered counterpart~\cite{apf:96}
(for Method 1).

{\bf Method 1.}

Introducing the formal inner product 
\beq 
\left\langle \Phi , \Psi
\right\rangle (t) = \int_{ T^{d}} d^{d} {\bf x} \; \Phi (t,{\bf x}) \,
\Psi (t,{\bf x}) \; ,
\label{inner_product_continuous}
\eeq 
the action~(\ref{eq:action_continuous}) becomes 
\beq S [\Phi] =
 \frac{1}{2} 
\int_{ \Re^{1}} d t \; \left[ \left\langle \dot{\Phi} ,
 \dot{\Phi} \right\rangle (t) - 
\left\langle \mbox{\boldmath $\nabla$}
 \Phi , \mbox{\boldmath $\nabla$} \Phi \right\rangle (t) 
- m^{2}
\left\langle
 \Phi , \Phi \right\rangle (t) 
\right] 
\;   ,
\label{eq:action_continuous3}
\eeq 
which may be converted into an effective quantum-mechanical
problem by expanding the scalar field in periodic eigenfunctions 
\beq
\Phi(t,{\bf x}) = \sum_{{\bf n} \in Z^{d}} \phi^{ {\bf n} }(t) 
\, b_{{\bf n}}({\bf x})
\label{eq:KK_Fourier_continuous}
\;
\eeq
 (Kaluza-Klein-like decomposition)
and integrating out the ${\bf x}$ dependence.
In Eq.~(\ref{eq:KK_Fourier_continuous}),
$\left\{ b_{{\bf n}}({\bf x}) \right\}_{{\bf n} \in Z^{d}}$ is a basis 
for the space $\Re^{T^{d}}$
of real functions on the $d$-torus
(${\bf x} \in T^{d}$).
For the sake of simplicity and without loss of generality, we will
take the spatial coordinates as defined in $[-L/2,L/2]^{d}$,
with periodic boundary conditions;
even though it is customary to use the Fourier basis 
$b_{{\bf n}}({\bf x})=   
e^{2 \pi i {\bf n} \cdot {\bf x} /L} $, our analysis
will be carried out for an arbitrary 
$\left\{ b_{{\bf n}}({\bf x}) \right\}$.

In order to avoid the appearance of awkward divergences, we will 
work with the discrete version of the theory,
as defined in a Minkowskian spacetime lattice with 
compactified spatial coordinates
$(t_{\alpha},{\bf x}_{ {\bf j} })$.
Specifically,
the introduction of the 
large integers $M$ and $N$, as well as of a finite time interval $T$, 
defines the lattice spacings 
$ \delta=T/M$ and $ \epsilon =L/N$,
in terms of which
$t_{\alpha}= \alpha \, \delta$
and ${\bf x}_{ {\bf j} }= {\bf j} \,  \epsilon$,
with
$\alpha$ and ${\bf j}$ selected from the integers modulo 
$M$ and $N$ respectively,
i.e., $\alpha \in Z_{M}$ and
${\bf j} = (j_{1}, \ldots , j_{d})
 \in (Z_{N})^{d}$.
In what follows, it will prove useful to introduce the notation
$\varphi^{ {\bf j} }(t) = 
\Phi (t,{\bf x}_{ {\bf j} })$, 
with which the lattice action becomes
\begin{equation}
S [\varphi]
= \frac{1}{2}  \,
\delta  \, \epsilon^{d}
\sum_{\alpha \in Z_{M}} \, 
\sum_{ {\bf j}  \in (Z_{N})^{d} }
\Biggl\{ 
\left[
\frac{
\varphi^{ {\bf j} }(t_{\alpha +1})-\varphi^{ {\bf j} }
(t_{\alpha}) }{\delta }
 \right]^2 
 - 
 \sum_{\mu=1}^{d} 
\left[ 
\frac{ \varphi^{ {\bf j} + {\bf e}_{\mu}
}(t_{\alpha})-\varphi^{ {\bf j} }(t_{\alpha})}{\epsilon }
\right]^2
 - 
m^{2}
\left[    \varphi^{ {\bf j} }  (t_{\alpha}) 
 \right]^2 
\Biggr\} 
\;  ,
\label{eq:action_discrete}
\end{equation}
where ${\bf e}_{\mu}$ is the unit vector in
the $\mu$ direction.
In this Letter,
we will focus on the discretization of the spatial variable,
as a way of introducing a quantum-mechanical system with a finite number
of degrees of freedom.
Instead, the time variable will be kept continuous in most equations,
with the understanding that discretization of the 
time---independently from ${\bf x}$---can be implemented whenever 
this proves convenient.\footnote{In Ref.~\cite{apf:96}, 
it was shown that the use of a lattice
for the variable $t$ (with an independent lattice constant $\delta $
arising from an even number $M$ of points)
 permits the correct quantum-mechanical treatment of ``extra'' terms
$O(\hbar^{2})$ in the limit $M \rightarrow \infty$,
under a nonlinear change of variables.}

The main advantage of introducing a lattice
for our problem lies in that it eliminates ultraviolet divergences,
by reducing the space of functions defined on 
$T^{d}$ to a finite-dimensional space 
$ \Re^{(Z_{N})^{d}} \equiv
{\cal V}_{N^{d}}$, of dimension $N^{d}$.
In ${\cal V}_{N^{d}}$,
an arbitrary basis
can be chosen by selecting $N^{d}$ linearly independent vectors 
$ \mbox{\boldmath\large  $ \left\{  \right.$ } \! \!
  \mbox{\boldmath\large  $ \left(  \right.$ } \! \!
b_{{\bf n}}({\bf x}_{{\bf j}}) 
\!   \mbox{\boldmath\large  $ \left.  \right)$}  
_{
{\bf j} \in (Z_{N})^{d}}
\!   \mbox{\boldmath\large  $ \left.  \right\}$}  
_{
{\bf n} \in (Z_{N})^{d}}
$.
Then, for any field variable,
\beq
\Phi(t,{\bf x}) = \sum_{ {\bf n} 
 \in (Z_{N})^{d}} \phi^{{\bf n}}(t) 
\,
b_{{\bf n}}({\bf x})
\label{eq:KK_Fourier}
\; ,
\eeq
which reproduces the Kaluza-Klein 
decomposition~(\ref{eq:KK_Fourier_continuous}) in the continuum limit,
while
\beq
\varphi^{ {\bf j} }(t) = \sum_{{\bf n} \in (Z_{N})^{d}} \phi^{{\bf n}}(t) 
\,
\Lambda^{ {\bf j } }_{{\bf n}}
\label{eq:KK_Fourier_matrix}
\; ,
\eeq
with $\Lambda^{ {\bf j} }_{{\bf n}} =
 b_{{\bf n}} ({\bf x}_{{\bf j}})$ defining an invertible 
$N^{d} \times N^{d}$ matrix.
A particular convenient choice, in addition to the Fourier basis,
is provided by the ``canonical''
basis $c_{ {\bf j} }({\bf x})$, which is defined by 
$c_{ {\bf j} }({\bf x}_{\bf k})=\delta^{ {\bf k}}_{ {\bf j} }$
and amounts to a real-space lattice representation of the field 
$\Phi(t,{\bf x})$,
with components $\varphi^{ {\bf j} }(t)$.
The discrete version of the inner 
product~(\ref{inner_product_continuous}),
\beq
\left\langle \Phi , \Psi  \right\rangle (t)
= \left( \frac{L}{N} \right)^{d}
\, \sum_{ {\bf j} \in (Z_{N})^{d}}
\varphi^{ {\bf j } } (t) \,
\psi^{{\bf j} } (t) 
= \sum_{{\bf n},{\bf m} \, \in (Z_{N})^{d} }
\gamma_{{\bf n}{\bf m}}
\,
 \phi^{{\bf n}} (t) 
\psi^{{\bf m}} (t) 
\;  
\label{inner_product}
\eeq
defines the linear-space symmetric metric
$
\gamma_{{\bf n}{\bf m}}
=\left\langle b_{{\bf n}} , b_{{\bf m}} \right\rangle
$,
 in terms of which the resulting action is
\beq
S [\phi]  
= 
\int d t 
\sum_{{\bf n},{\bf m} \in (Z_{N})^{d}} 
\left[ 
\half g_{{\bf n}{\bf m}}[\phi] \dot{\phi}^{{\bf n}} \dot{\phi}^{{\bf m}}
- \half  h_{{\bf n}{\bf m}}  \phi^{{\bf n}} \phi^{{\bf m}} 
- \frac{ m^{2} }{2} \,
 \gamma_{{\bf n}{\bf m}}
 \phi^{{\bf n}} \phi^{{\bf m}}
\right]
\; 
\label{eq:action_discrete_KK} 
\eeq
[cf.\ Eq.~(\ref{eq:action_continuous3})],
where the matrix elements 
$g_{{\bf n}{\bf m}}  [\phi]  $ (metric) and $h_{{\bf n}{\bf m}}  $,
and $ \gamma_{{\bf n}{\bf m}}$
admit the expressions 
\beq
g[\phi] \equiv 
\left\langle 
            \frac{\partial \varphi}{\partial \phi^{{\bf n}} } 
,
            \frac{\partial \varphi}{\partial \phi^{{\bf m}} } 
\right\rangle
= 
\left\langle b_{{\bf n}} , b_{{\bf m}} \right\rangle
\equiv
  \gamma
\; \; \; \; \; 
;
\; \;
h  =  \sum_{\mu=1}^{d}
\left( \nabla_{\! \! \mu}  \right)^{T}
 \gamma  \,
\nabla_{\! \! \mu}
\;  .
\label{eq:metrics}
\eeq 
In Eq.~(\ref{eq:metrics}) the elements of the matrix 
$\nabla_{\! \! \mu}$ are defined in terms of 
the lattice counterparts of the spatial derivatives 
$\partial_{\mu} b_{{\bf n}} ({\bf x})$, i.e., 
\begin{equation}
b_{{\bf n}}( {\bf x}_{ {\bf j} + 
{\bf e}_{\mu} } ) -
b_{{\bf n}}({\bf x}_{{\bf j}}) 
= \epsilon {\displaystyle \sum_{ {\bf l} \in (Z_{N})^{d}} } \, 
( \nabla_{\! \! \mu} )^{ {\bf l} }_{{\bf n}} 
\, b_{ {\bf l}} ({\bf x}_{{\bf j}})
\; .
\label{eq:discrete_gradient}
\end{equation}  
With this definition,
the matrix $\nabla_{\! \! \mu}$ is explicitly dependent on $\epsilon$
or $1/N$, i.e., $\nabla_{\! \! \mu}= \nabla_{\! \! \mu}(1/N)$; 
however, it admits the
asymptotic expansion 
$ \nabla_{\! \! \mu}(1/N)= \nabla_{\! \! \mu}(0) +O(1/N)$, so that a
definite finite value $\nabla_{\! \! \mu}^{(0)}= \nabla_{\! \! \mu} (0)$ 
exists in the
continuum limit ($N \rightarrow \infty$). This limit will be
eventually assumed in Eq.~(\ref{eq:action_discrete_KK}) and similar
expressions, in which case the substitution $(Z_{N})^{d} \rightarrow
Z^{d}$ should be performed.\footnote{Parenthetically,
Eq.~(\ref{eq:action_discrete_KK}) describes a system of $N$ coupled
quantum-mechanical oscillators $\phi^{{\bf n}}(t) $; for example, when
the Fourier basis 
$b_{{\bf n}}({\bf x})=e^{2 \pi i {\bf n} \cdot {\bf x}/L } $ 
is chosen, then 
$\gamma_{{\bf n}{\bf m}}= L^{d} \,
\delta_{{\bf n},-{\bf m}} $ 
and 
$
 \mbox{\boldmath  $ \nabla$}^{{\bf n}}_{{\bf m}} =
 2 \pi i \, {\bf n} \, \delta^{{\bf n}}_{{\bf m}}  /L 
+
O(1/N)$, 
whence Eq.~(\ref{eq:action_discrete_KK}) provides the
frequencies $\omega_{\bf n} = 
\sqrt{ (2 \pi | {\bf n} | /L)^{2}+m^{2}} $, as $N \rightarrow
\infty$.}

Let us now consider the field redefinition~(\ref{eq:Ntran}) and
expand the new field $\tilde{\Phi} (t,{\bf x})$ in modes,    
\beq
\tilde{\Phi}(t,{\bf x}) = \sum_{{\bf n} \in (Z_{N})^{d} } 
\tilde{\phi}^{{\bf n}} (t)  \,  
b_{ {\bf n}} ( {\bf x} ) 
\; ,
\eeq
with the implicit transformation
\beq
\phi^{{\bf n}} \equiv f^{{\bf n}} [\tilde{\phi}]  
\;  .
\label{eq:atob}
\eeq
Then, the reduced metric  $g[\tilde{\Phi}]$
 (with respect to the new coordinates), as defined
in Eq.~(\ref{eq:ultralocal_metric}) from the ultralocal full-fledged 
metric ${\cal G} [ \tilde{\Phi} ] (t,{\bf x};t^{\prime},{\bf x}^{\prime})$,
is diagonal; in fact, the lattice version of 
Eq.~(\ref{eq:ultralocal_metric}) 
implies that the reduced lattice metric coefficients 
$g_{\bf jk}[\tilde{\varphi}]$ 
are given from
\beq
g[\tilde{\varphi}]
= \left( \frac{L}{N} \right)^{d} 
\, {\rm diag}  
\left\{
\left(
 F^{\prime}[\tilde{\varphi}] 
\right)^{2}
\right\}
\;  .
\label{eq:real_space_metric}
\eeq
On the other hand,  with respect to any other basis,
\beq
g[\tilde{\phi}]
= \Lambda^{T}
g[\tilde{\varphi}]
\Lambda
\;  ,
\label{metric_transf}
\eeq
where 
$\Lambda^{ {\bf j}}_{{\bf n}} =
\partial \tilde{\varphi}^{ {\bf j} }/
\partial \tilde{\phi}^{{\bf n}}= b_{{\bf n}}({\bf x}_{{\bf j}})$.

The change of variables~(\ref{eq:atob})
in the quantum-mechanical path integral should be implemented 
by including the ``extra'' term, i.e.,
the transformed action $S[\tilde{\phi}]$ becomes
\beq 
S[\tilde{\phi}]=
S_{0}[\tilde{\phi}] + 
S_{\rm Jacobian}[\tilde{\phi}]
 + S_{\rm extra} [\tilde{\phi}]
\; .
\label{eq:transformed_action}
\eeq 
The first term in Eq.~(\ref{eq:transformed_action}) can be
computed by direct substitution in Eq.~(\ref{eq:action_discrete_KK}),
\beq
S_{0}[\tilde{\phi}]  
=
S \left[ f[\tilde{\phi}] \right]  
= 
\int d t \;
   \sum_{{\bf n},{\bf m} \in (Z_{N})^{d}} 
 \left[ 
\half \,
g_{{\bf n}{\bf m}}  [\tilde{\phi}]
 \dot{\tilde{\phi^{{\bf n}}}}
 \dot{\tilde{\phi^{{\bf m}}}}
-  \half \,
\left(
 h_{{\bf n}{\bf m}} 
+ m^{2} \gamma_{{\bf n}{\bf m}}  
\right) 
 f^{{\bf n}} [\tilde{\phi}] f^{{\bf m}}[\tilde{\phi}] \right]
\; .
\label{eq:m10}
\eeq
As for the second term,
the Jacobian determinant
${\displaystyle \prod_{\alpha \in Z_{M} } } 
\left\{ \left( \det \check{g}
 [\tilde{\phi}] 
(t_{\alpha}) \right)^{1/2} \right\}$---with $\check{g} [\tilde{\phi}] 
= g  [\tilde{\phi}]  \, \delta$
[from Eq.~(\ref{eq:ultralocal_metric})]---leads 
to the standard contribution to the action,
\beq
S_{\rm Jacobian}[\tilde{\phi}] 
=
-\frac{ i \hbar}{2}  \,
\sum_{\alpha \in Z_{M}} 
{\rm tr} \,
\ln   \left\{
\check{g} [\tilde{\phi}] (t_{\alpha})
\right\}
=
-\frac{ i \hbar}{2}  
\,
{\rm Tr} \, \ln 
\left\{
g [\tilde{\phi}] \delta (t -t^\prime )
\right\}
\; 
\label{eq:m1j}
\eeq
[cf.\ Eq.~(\ref{eq:jacobian_spacetime})], 
where
${\rm tr}$ stands for the reduced spatial trace (with respect
to spatial indices alone), as opposed to the spacetime trace ${\rm Tr}$.
Finally, the ``extra'' term 
arising from the stochastic 
nature of the path integral is\footnote{The 
Einstein summation convention for repeated indices is adopted from 
Eq.~(\ref{eq:action_extra_QM}) on.}
\beq
S_{\rm extra} [\tilde{\phi}] 
= - \frac{\hbar^2}{8} \int d t \; 
 g^{{\bf n}{\bf m}} [\tilde{\phi}] \, 
\Gamma^{ {\bf s}}_{{\bf l}{\bf n}}[\tilde{\phi}]  \,
\Gamma^{{\bf l}}_{{\bf s}{\bf m}}[\tilde{\phi}]  
= 
- \frac{\hbar^2}{8} \int d t \; 
 {\rm tr} \, \left( g^{-1}[\tilde{\phi}] 
\, \Xi [\tilde{\phi}] \right)
\; ,
\label{eq:action_extra_QM}
\eeq
where 
$\Gamma^{{\bf s}}_{{\bf l}{\bf n}} [\tilde{\phi}]  $ 
are the connection coefficients 
associated with the metric $g_{{\bf n}{\bf m}} [\tilde{\phi}]  $ and
 $\Xi_{{\bf n}{\bf m}}= 
\Gamma^{{\bf s}}_{{\bf l}{\bf n}} \,  \Gamma^{{\bf l}}_{{\bf s}{\bf m}}$.
Equation~(\ref{eq:action_extra_QM})
requires the evaluation
of ${\rm tr} \, \left( g^{-1}[\tilde{\phi}] \, 
\Xi [\tilde{\phi}] \right)$,
which can be performed in an arbitrary basis
$\left\{ b_{{\bf n}} \right\}$,
due to the tensor nature of the expressions involved in the
lattice version of the theory.
However, this is most easily done in real space,
where the metric is diagonal\footnote{Obviously, when the metric
is nondiagonal, the computations are quite a bit lengthier.
For example,
for  the Fourier basis of exponentials
and $\Phi = \tilde{\Phi} +
  \lambda \tilde{\Phi}^{\nu} $,
 the same results follow straightforwardly from
$g [\tilde{\phi}]
= (L/ N )^{d}
[ 1 + \lambda \nu (\tilde{\Phi})^{\nu-1}]^{2}$,
with the matrix 
$(\tilde{\Phi})^{{\bf n}}_{{\bf m}} 
= \tilde{\phi}^{{\bf n}-{\bf m}}$.}
[Eq.~(\ref{eq:real_space_metric})],
so that the inverse metric 
is $g^{-1}[\tilde{\varphi}]= 
( L/N )^{-d} 
 \, {\rm diag}
\left\{  \left( F^{\prime}[\tilde{\varphi}] \right)^{-2}
\right\}
$
and the connection coefficients are
$\Gamma [\tilde{\varphi}]= 
 {\rm diag}^{(3)}
\left\{   F^{\prime \prime }[\tilde{\varphi}] /
F^{\prime }[\tilde{\varphi}] 
\right\}
$ (diagonal with respect to the three indices in real space).
Then,
\beq
 {\rm tr} \, \left( g^{-1}[\tilde{\varphi}] 
\, \Xi [\tilde{\varphi}] \right)
=
\left( \frac{L}{N} \right)^{-d} 
 \,
 \sum_{ {\bf j} \in (Z_{N})^{d}} 
\left[
\frac{ \left( F^{\prime \prime} [\tilde{\varphi}]\right)^{2} }{ 
\left( F^{\prime}[\tilde{\varphi}] \right)^{4} }
\right]_{\bf j}
\;  .
\label{eq:VE_discrete}
\eeq

For later comparison with Method 2, it is useful to rewrite
Eqs.~(\ref{eq:action_extra_QM}) and (\ref{eq:VE_discrete}) 
explicitly in terms of the field $\tilde{\Phi} $; then,
\beq
S_{\rm extra}[\tilde{\Phi}] 
= 
-\frac{\hbar^2}{8} 
\left( \frac{L}{N} \right)^{-2d} 
\int d^{d+1}  x \;
\frac{ \left( F^{\prime \prime} [\tilde{\Phi}]\right)^{2} }{ 
\left( F^{\prime}[\tilde{\Phi}] \right)^{4} }
\; .
\label{eq:VE}
\eeq

{\bf Method 2.}

In this method, the field redefinition~(\ref{eq:Ntran})
is applied first, while the expansion in Kaluza-Klein modes is later 
performed in the transformed field theory.
The terms in the action obtained
from Method 2 will be written with hats to distinguish them from those of 
Method 1. 
By direct substitution of the field redefinition~(\ref{eq:Ntran}) in
the action~(\ref{eq:action_continuous}),
the piece
\beq 
\widehat{S}_{0}[\tilde{\Phi}] 
= 
S \left[ F[ \tilde{\Phi}] \right] 
= 
\half \int d^{d+1} x  \; 
\left[
\eta^{\mu \nu} 
\left( F^{\prime}[\tilde{\Phi}] \right)^{2}
\partial_\mu \tilde{\Phi} \,  \partial_\nu \tilde{\Phi}  
- m^{2} \left( F [\tilde{\Phi}] \right)^{2}
\right]
\; 
\eeq 
develops derivative interaction terms,
while the Jacobian, from 
Eqs.~(\ref{eq:ultralocal_metric}) and
(\ref{eq:jacobian_spacetime}), 
yields
\beq 
\widehat{S}_{\rm Jacobian}
[\tilde{\Phi}] 
=
 - i \hbar \delta^{(d+1)}(0) \int  d^{d+1}  x
\;
{\rm ln} \, F^{\prime}[\tilde{\Phi}]
\;  .
\label{eq:m2j}
\eeq
The conventional arguments within the standard lore
would imply that the total action is given by
only these two contributions,  
$ \widehat{S} [\tilde{\Phi}] =
\widehat{S}_{0} [\tilde{\Phi}] + 
\widehat{S}_{\rm Jacobian}[\tilde{\Phi}].$  

Finally,
the action $\widehat{S} [\tilde{\Phi}] $
can be converted into an effective quantum-mechanical one, 
$\widehat{S} [\tilde{\phi}] = 
\widehat{S}_{0} [\tilde{\phi}]
+
\widehat{S}_{\rm Jacobian}[\tilde{\phi}] $, by
using Eq.~(\ref{eq:KK_Fourier_continuous}) (for
$\tilde{\Phi}$) and
integrating out the spatial coordinates, with the results
\beq 
\widehat{S}_{0}
[\tilde{\phi}]
 =  \int d t \; 
\sum_{{\bf n},{\bf m} \in Z} 
\left[ \half  \widehat{g}_{{\bf n}{\bf m}}[\tilde{\phi}]  
\dot{\tilde{\phi^{{\bf n}} }} \dot{\tilde{\phi^{{\bf m}} }}
-
\half 
\left(
\widehat{h}_{{\bf n}{\bf m}}
+ m^{2} \widehat{\gamma}_{{\bf n}{\bf m}}  
\right)
 f^{{\bf n}} [\tilde{\phi}] 
f^{{\bf m}}[\tilde{\phi}] 
\right] 
\;  
\label{eq:m20}
\eeq 
and [from Eqs.~(\ref{eq:ultralocal_metric})
and (\ref{eq:jacobian_spacetime})]
\beq
\widehat{S}_{\rm Jacobian}[\tilde{\phi}] 
=
-\frac{ i \hbar}{2}  
\; {\rm Tr} \ \ln 
\left\{ 
\widehat{g} [\tilde{\phi}] \delta (t -t^\prime )
\right\}
\; .
\label{eq:m2j_reduced}
\eeq
In Eqs.~(\ref{eq:m20}) and (\ref{eq:m2j_reduced}),
\begin{equation}
\widehat{g}_{{\bf n}{\bf m}} [\tilde{\phi}]
\equiv
\int d^{d} {\bf x} \;
 b_{{\bf n}}({\bf x}) 
\,
b_{{\bf m}}({\bf x}) 
\left( F^{\prime}[\tilde{\Phi} (t,{\bf x})] \right)^{2}
=
g_{{\bf n}{\bf m}} [\tilde{\phi}]
\; , 
\end{equation}
as follows from the limit $N \rightarrow \infty$ of
Eqs.~(\ref{eq:real_space_metric}) and 
(\ref{metric_transf});
likewise
$\widehat{h}_{{\bf n}{\bf m}} = {h}_{{\bf n}{\bf m}}$
[from Eqs.~(\ref{eq:metrics}) and (\ref{eq:discrete_gradient})]
and
$\widehat{\gamma}_{{\bf n}{\bf m}} = {\gamma}_{{\bf n}{\bf m}}$,
so that  
$\widehat{g}_{{\bf n}{\bf m}} [\tilde{\phi}]$,
$\widehat{h}_{{\bf n}{\bf m}} $,
and $\widehat{\gamma}_{{\bf n}{\bf m}} $
 coincide with the corresponding
matrix elements appearing in the 
continuum limit of the quantum-mechanical
version of this calculation.

{\bf Comparison of Methods.}

The transformations involved in Methods 1 and 2 are 
represented in the diagram
\begin{equation}
\diagram{ S[\Phi] 
  &  \diagramrightarrow{ \mbox{$  \Phi = F[\tilde{\Phi}] $ }  }{}    
          & 
  \widehat{S} [\tilde{\Phi}]
 \cr
 \diagramdownarrow{\cal R }{}  & & 
  \diagramdownarrow{\cal R }{}   
                                       \cr
\diagram{ S[\phi] }
  &
\diagramrightarrow{  \mbox{$  \phi = f[\tilde{\phi}] $ }  }{}    
    &   
 S
[\tilde{\phi}]  
\stackrel{?}{=} 
 \widehat{ S } [\tilde{\phi}]  
 \cr
}
\;  \;  ,
\label{eq:diagram}
\end{equation}
where ${\cal R}$ stands for reduction
to a quantum-mechanical problem
(by integrating out the spatial coordinates).
The identity of the results of the two methods 
amounts to the 
equality of the two actions for the effective quantum-mechanical theory;
in other words, it is equivalent to the statement that
diagram~(\ref{eq:diagram}) be commutative.
However, if the standard lore holds true, 
the action $ \widehat{ S } [\tilde{\phi}]  $
lacks the ``extra'' term, so that
\beq 
S_{0}[\tilde{\phi}] 
+ S_{\rm Jacobian}[\tilde{\phi}] 
+ S_{\rm extra} [\tilde{\phi}] = 
\widehat{S}_{0}[\tilde{\phi}] + 
\widehat{S}_{\rm Jacobian}[\tilde{\phi}]  
\label{eq:action_comparison}
\; .
\eeq 

Let us now analyze the feasibility of Eq.~(\ref{eq:action_comparison}).
Firstly,
the equality $S_{0}[\tilde{\phi}] =
\widehat{S}_{0}[\tilde{\phi}] $ 
follows from Eqs.~(\ref{eq:m10})
and (\ref{eq:m20}).
Secondly, the equality of the Jacobian factors,
 $ S_{\rm Jacobian}[\tilde{\phi}] 
 = \widehat{S}_{\rm Jacobian}[\tilde{\phi}]  $
is seen from
Eqs.~(\ref{eq:m1j}) and~(\ref{eq:m2j_reduced}).
Finally, 
due to the identity of the first two terms,
it is clear that Eq.~(\ref{eq:action_comparison})
is incompatible with the existence of a nonzero term
$S_{\rm extra}[\tilde{\phi}]$.  
In other words, we are now confronted with
the central issue of this Letter:
in Method 1, $S_{\rm extra}[\tilde{\phi}] \neq 0$, while in 
Method 2, the standard rules for
nonlinear field redefinitions failed to generate such a term.
The inescapable conclusion, if $S_{\rm extra}[\tilde{\phi}] $
cannot be rationalized to vanish, is that
this term should have emerged at the level of quantum field
theory from the nonlinear field redefinition.
Therefore, from Eq.~(\ref{eq:VE})
and the identification
\beq
\left( \frac{L}{N} \right)^{-d} 
   =  \delta^{(d)} ({\bf x}=0)
\label{eq:sumtodelta}
\;  
\eeq
(in the limit $N \rightarrow \infty$)---which 
is recognized to be the standard condition for the transition
from the lattice version of the theory to its continuous counterpart---it
follows that
the final expression for the ``extra'' quantum-field theoretical term is 
\beq
S_{\rm extra}[\tilde{\Phi}] = -\frac{\hbar^2}{8} 
\left[ \delta^{(d)}({\bf x}=0) 
\right]^2
\int
d^{d+1} x \;
\frac{ \left( F^{\prime \prime} [\tilde{\Phi}]\right)^{2} }{ 
\left( F^{\prime}[\tilde{\Phi}] \right)^{4} }
\; .
\label{eq:kkextra}
\eeq
This divergent term is proportional to the
square of the $d$-dimensional spatial delta function rather than 
the $(d+1)$-dimensional delta function at zero argument.  
An analogue of this result was found 
in the early literature on four-dimensional
chiral dynamics~\cite{dow:71,suz:72,cha:73}.

A final remark is in order.
An alternative to the approach of 
Method 2 is afforded by the addition, after 
field redefinition,
of an infinite series of counterterms,
\begin{eqnarray}
S[\tilde{\Phi}]  = 
\half \int d^{d+1} x 
\;
\Biggl[
& \eta^{\mu \nu} &
\left( F^{\prime}[\tilde{\Phi}] \right)^{2}
\partial_\mu \tilde{\Phi} \,  \partial_\nu \tilde{\Phi}  
- m^{2} \left( F [\tilde{\Phi}] \right)^{2}
\Biggr]
\nonumber \\
 & - &
i \hbar \delta^{(d+1)}(0) \int d^{d+1} x \;
{\rm ln} \, F^{\prime}[\tilde{\Phi}]
+ \int d^{d+1} x \;
 \sum_{\ell=1}^\infty c_\ell \, 
\tilde{\Phi}^\ell 
\;  ,
\label{eq:counterterms}
\end{eqnarray}
 where the unknown coefficients $c_\ell$  
can be evaluated by computing physically significant 
quantities and matching with the original free theory.   
The advantage of our approach lies in that
we have been able to directly derive the simple expression in 
Eq.~(\ref{eq:kkextra}),
which would otherwise be obtained by laboriously computing
Feynman diagrams and summing the series in Eq.~(\ref{eq:counterterms}).

In conclusion,
we have shown evidence for a single ``extra'' term being generated 
upon making nonlinear field redefinitions for the (d+1)-dimensional 
quantum field theory in a Minkowskian spacetime
with compactified spatial coordinates.
An extension of the work in quantum 
mechanics~\cite{apf:96},
as well as a perturbative calculation based upon these results,
gives additional confirmation of the existence of ``extra'' terms in 
quantum field theory, at least for the case of flat Euclidean 
$D$-dimensional spacetime. Finally, this 
work also reveals the need for a more careful use of dimensional 
regularization in higher-order calculations, as will be discussed
elsewhere.

{\bf Acknowledgements.}

C.R.O.'s work was supported in part by
an Advanced Research Grant from the Texas
Higher Education Coordinating Board.
H.E.C. acknowledges financial support by
the University of San Francisco Faculty Development Fund,
as well as the hospitality of the University of Houston.

\end{document}